\documentstyle[psfig]{mn}

\newcommand{\ltapprox}{\raisebox{-0.5ex}{$\,\stackrel{<}{\scriptstyle 
\sim}\,$}}
\newcommand{\gtapprox}{\raisebox{-0.5ex}{$\,\stackrel{>}{\scriptstyle 
\sim}\,$}}

\title[Spectroscopic Confirmation of Redshifts Predicted by Gravitational
	Lensing]
	{Spectroscopic Confirmation of Redshifts Predicted by Gravitational
	Lensing
	\thanks{Based on observations obtained on the William Herschel
	Telescope at the Observatorio del Roque de los Muchachos, La Palma.}}

\author[T.M.D. Ebbels et. al.]
       {Tim Ebbels$^{1}$, Richard Ellis$^{1}$, Jean-Paul Kneib$^{1}$,\cr
        Jean-Fran\c cois Le Borgne$^{2}$, Roser Pell\'o$^{2}$,
	Ian Smail$^{3}$ \& Blai Sanahuja$^{4}$\\
$^{1}$Institute of Astronomy, Madingley Road, Cambridge CB3 0HA, U.K.\\
$^{2}$Observatoire Midi-Pyr\'en\'ees, 14 Av. E.Belin, 31400 Toulouse, France.\\
$^{3}$Department of Physics, University of Durham, South Road,
       Durham. DH1 3LE\\
$^{4}$Departament d'Astronomia i Meteorologia, Universitat de Barcelona,
Diagonal 648, 08028 Barcelona, Spain.\\}

\date{Accepted 1997  
      Received 1997 ;
      in original form }

\begin{document}

\maketitle

\begin{abstract}

We present deep spectroscopic measurements of 18 distant field
galaxies identified as gravitationally-lensed arcs in a {\it Hubble
Space Telescope} image of the cluster Abell 2218. Redshifts of these
objects were predicted by Kneib et al.\ (1996) using a lensing
analysis constrained by the properties of two bright arcs of known
redshift and other multiply-imaged sources. The new spectroscopic
identifications were obtained using long exposures with the LDSS-2
spectrograph on the William Herschel Telescope and demonstrate the
capability of that instrument to new limits, $R\simeq$24; the lensing
magnification implies true source magnitudes as faint as
$R\simeq$25. Statistically, our measured redshifts are in excellent
agreement with those predicted from Kneib et al.'s lensing analysis
which gives considerable support to the redshift distribution derived
by the lensing inversion method for the more numerous and fainter
arclets extending to $R\simeq 25.5$. We explore the remaining
uncertainties arising from both the mass distribution in the central
regions of Abell 2218 and the inversion method itself, and conclude
that the mean redshift of the faint field population at $R\simeq$25.5
($B\sim 26$--27) is low, $<\! z\! >$=0.8--1.  We discuss this result
in the context of redshift distributions estimated from multi-colour
photometry. Although such comparisons are not straightforward, we
suggest that photometric techniques may achieve a reasonable level of
agreement particularly when they include near-infrared photometry with
discriminatory capabilities in the $1<z<$2 range.

\end{abstract}


\section{Introduction}

Gravitational lensing is now established as a highly-successful
technique for studying both the mass distribution on various scales in
rich clusters at intermediate depth, and for constraining the properties
and distances of very faint sources magnified serendipitously via the
lensing cluster. Smail et al.\ (1994, 1995a) attempted to break the
degeneracy between these two important applications by studying the
lensing signal in a number of rich clusters chosen to lie at different
redshifts. Although they concluded the bulk of a faint galaxy sample
limited at $I\simeq$25 has a redshift distribution with a median close
to one, their conclusion depended somewhat on the mass distribution in the
more distant and hence less well-understood clusters.

The discovery of multiply-imaged systems in the cores of rich clusters
(Mellier, Fort \& Kneib 1993, Kneib et al.\ 1993, Smail et al.\ 1995b)
offers an alternative and more promising route. The mass models for
clusters containing such multiple images are sufficiently constrained
by the location and orientation of the images that it is possible to
use these models to match the predicted shear of the fainter images
with their observed shapes directly. In the case of multiple images,
the requirement that each image must come from the same object in the
source plane provides many constraints leading to a rather precise
redshift prediction. Singly imaged arclets generate fewer constraints
but nonetheless, a most likely redshift may be determined for each arclet, and
useful statistical results can be obtained for the background
redshift distribution, provided the sample size is large enough.  A
first application of this so-called `lens inversion' technique was
discussed by Kneib et al.\ (1994) for the cluster Abell 370. Individual
redshifts were obtained for 30 candidate arclets limited at $B$=27
with axial ratios $a/b>$1.4 measured on ground-based images.

The inversion technique was considered in more detail by Kneib et al
(1996, hereafter KESCS) on the basis of an impressive image of Abell
2218 ($z$=0.175) obtained with the refurbished {\it Hubble Space Telescope}
(HST), Figure \ref{fig-hstimage}. The important role of HST
imaging in lensing studies was demonstrated in two respects. Firstly,
via the superlative resolution of HST, the multiply imaged nature of
several of the most strongly lensed arcs was revealed in spectacular
detail. Secondly, in the absence of ground-based seeing, the fainter
sheared images were individually identified with much greater
confidence than was possible using ground-based data. This, in turn,
enabled accurate inversion for a greater number of faint sources.
Predicted redshifts were determined by KESCS for $\simeq$80 arclets in
Abell 2218 to $R$=25 and a formalism was developed to predict the
uncertainties arising in the inversion from a number of sources. The
limitations of lensing inversion as a general surveying technique were
also discussed.

Although impressive progress has been made in locating high redshift
galaxies via deep ($K<$20) magnitude-limited redshift surveys 
(Cowie et al.\ 1995, 1996) and via various colour selection criteria such 
as those sensitive to the presence of a Lyman limit in the near ultraviolet
(Steidel et al.\ 1996), an important advantage of lensing inversion as a 
purely geometric technique over other methods is that it does not require 
the presence of a particular spectral feature within some restricted observational passband. This suggests it should complement the other 
techniques by providing an accurate mean redshift distribution at 
various limits even if, ultimately, inversion through several clusters 
at various redshifts may be required to secure robust results.

Given the promise of the method it is clearly important to investigate the
uncertainties of lensing inversion by direct spectroscopic measurement
of the brighter arclets. This paper is concerned with executing such a
test for Abell 2218 whose mass model is highly constrained from 7
multiple image sets, two of which already have spectroscopic redshifts (Pell\'o
et al.\ 1992). Whereas only a few of the brighter arclets are amenable to
direct spectroscopy with current facilities, the test should still be
valuable since inversion is primarily a geometric technique. The
estimated redshift for a particular arclet relies on the cluster mass
distribution, the relative distances of the source and lens, and
uncertainties in the source and image shape (whose contribution can be
readily accounted for --  see KESCS). Thus a comparison of the
spectroscopic and predicted redshifts for the brighter arclets can be
made as valid a test of the method as that, more difficult, test applied
to the entire sample of arclets.

We have begun a survey of those arclets amenable for spectroscopic
study with the 4.2m William Herschel Telescope (WHT). Our aim is to
verify the inversion technique and, if necessary, correct the
residuals and produce an improved mass model. This in turn will lead
to better inversion for the more numerous and fainter arclets. As part
of this programme, one arclet in the sample (\#384 in the numbering
scheme of Le Borgne et al.\ 1992) has already been discussed by Ebbels
et al.\ (1996) owing to its high redshift ($z$=2.515) and the
spectacular agreement with the lensing prediction
($z=2.8^{+0.5}_{-0.3}$), making it an object of interest in its own
right.  A plan of the paper follows. In \S2 we discuss the selection
criteria for our sample of arclets in Abell 2218 and present the
journal of observations and spectroscopic results. In \S3 we compare
the observed redshifts with those predicted and consider the
uncertainties. In \S4 we review the mass model and illustrate those
changes which could still be permitted whilst being consistent with
the new sample of arclet redshifts.  Finally, in \S5 we return to the
question of the mean redshift of the faint population following the
discussion of KESCS.  \S6 summarizes our conclusions.

\section{Observations and Spectroscopic Redshifts}

The primary source material for our arclet sample is the HST WFPC2
image of A2218 as published by KESCS and shown in Figure
\ref{fig-hstimage}. The three orbit exposure totaling 6500s was taken
in the F702W passband and the magnitudes converted to standard R using
corrections from Holtzman et al.\ (1995). The resulting object
catalogue reaches $R=26$ with a completeness of 55\% at
$R\simeq25$. Ground based $(B-I)$ colours for our targets were measured
from images obtained with the COSMIC imaging spectrograph on the 5-m
Hale telescope at Palomar during June 1994 and June 1995. A thick
2048$^2$ TEK CCD ($0.284\ arcsec\ pix^{-1}$) was used and individual
exposures of $\sim 500$--1000s were taken using in--field dithering
(by $\sim 60$ arcsec). The images were reduced in a standard manner
with {\sc IRAF} using both twilight and on--source flat fields and
including cosmic ray rejection. Stacked exposure times totaled 16.5 ks
in $B$ and 21.7 ks in $I$, reaching $1\sigma $ limiting surface
brightnesses of $\mu_B = 28.2$ and $\mu_I = 26.8$ mag arcsec$^{-2}$,
while the final seeing on the frames measured 1.20 arcsec in $B$ and
0.95 arcsec in $I$. $B$ \& $I$ magnitudes were measured from these deep
exposures. Due to the faintness of these objects and the likelihood of
contamination by nearby bright cluster galaxies, we measured these
colours by hand. First, using IRAF's {\sc IMREPLACE} task, a
rectangular region enclosing the target object was replaced by fitting
a surface to the perimeter pixels and then using the distribution of
the perimeter pixel values to add sky noise to this surface. Then a
difference image was formed between the original and object-replaced
images such that only the flux from each object remained. Magnitudes
were measured within circular apertures of radius 1 or 2 arcseconds
and colours obtained from measurements within the same aperture in
each band.

In selecting our arclet sample from the HST image, we employed three
criteria. Firstly, we selected objects which looked distorted,
indicating that they might be gravitational arclets. As in KESCS, we
maximized the probability of selecting lensed images by restricting
the sample to those images whose orientation lies within $45\deg$ of
the local shear direction as predicted by the mass model. Secondly, to
maximize the probability of securing a reliable spectrum with
measurable features, we applied an integrated magnitude limit of
$B\simeq25$ (corresponding to $R\simeq24$) and gave highest priority
to objects bluer than $(B-I)\simeq2.2$. The former constraint
corresponds to the faintest objects within reach of the LDSS-2
spectrograph on the 4.2m WHT and the colour criterion is about half a
magnitude bluer than the cluster E/S0 sequence and increases the
probability of finding emission lines in the spectrum. Finally, we
were mindful that spectroscopic redshifts for some arclets would be
more important in the verification of the inversion method. The
uncertainty for each arclet is a function of location, shape and
redshift and thus by selecting those arclets for which the expected
error is small, we can provide a more stringent test of
inversion. These criteria were applied to 572 objects detected at 2
sigma above the sky in the WFPC2 image and produced a working
catalogue of some 65 candidate arclets.

\begin{figure*}

\caption{HST WFPC2 image of the cluster Abell 2218 in F702W showing
the sample of objects selected for spectroscopy (listed in table
\ref{tab-z}). Those for which we have obtained good redshifts are marked
with filled labels.
}
\label{fig-hstimage}
\end{figure*}

All observations were conducted with the LDSS-2 multi-object
spectrograph (Allington-Smith et al.\ 1994) on the 4.2 metre William
Herschel Telescope and are summarized in Table \ref{tab-obslog}. In
designing multi-slit masks, we were constrained by the fact that most
of the highly-magnified arcs lie close to the cluster core. This
caused problems with crowding since we insisted on a minimum slit
length of 10 arcseconds to enable good sky subtraction. A second
obvious limitation is the orientation and curvature of the
arclets. Although with ground-based seeing only the giant
arcs are noticeably non-linear in form, the orientations of the larger
arclets must be taken into account in order to maximize the amount of
light captured by each slit. Each mask had one primary target of this
type and thus we were able to base the mask orientation on the
inclination of this large arclet alone. 

\begin{figure}
\psfig{file=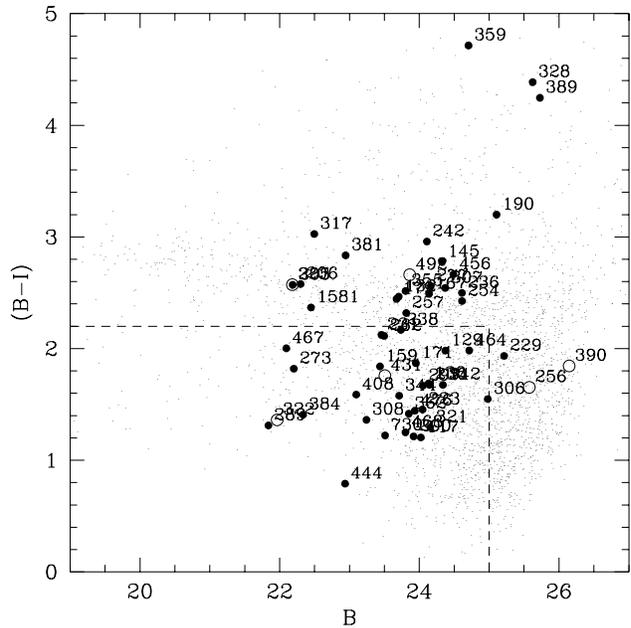,width=0.5\textwidth,angle=0}
\caption{The colour magnitude diagram for the arc sample superimposed
on that for the whole cluster field. Open circles denote non-sample
objects which were not obviously cluster members and the dashed lines
show our magnitude and colour criteria. Objects outside this region
were selected primarily by morphology or are serendipitous. Note the
similar $(B-I)$ colours of the multiple image systems (384,468) and
(359,328,389). }
\label{fig-BIB}
\end{figure}

Accordingly, from the above sample, 8 multi-slit masks were
constructed each containing between about 5 and 15 arclet targets which
allowed for some duplication between the two runs. In addition to the
main targets, spectra were obtained for several other objects, for
instance where a part of the slit happened to fall on a second object
near the main one or where a suitable arclet candidate was not
available. These `serendipitous' objects are at present included with
the main sample but flagged as such on the following diagrams. In summary
therefore, the spectroscopic sample consisted of 61 targets of which
37 were bona fide arclet candidates. The colour magnitude diagram for
our targets is presented in Figure \ref{fig-BIB}, superimposed on that
for the whole cluster field. Our magnitude and colour limits can
clearly be seen as can the similar (B-I) colours of several of the
multiple image systems. The observations were completed
during May/June 1995 and June 1996 under mixed conditions. A total of
7 masks were exposed as shown in Table \ref{tab-obslog}.

\begin{table*}
\centering
\caption{Spectroscopic Observations. The LDSS-2 medium blue grism was
used for all masks except mask 1 for which the medium red grism was
used. Both grisms give a resolution of $\sim12$\AA. During the 1995
run a thinned Tektronix 1024$^2$ CCD (TEK-1) was the detector, giving
$0.59\ arcsec\ pix^{-1} \& 5.3\AA pix{-1}$ while a new Loral 2048$^2$
CCD (LOR-1) became available for the 1996 run, with a scale of $0.37\
arcsec\ pix^{-1} \& 3.3 \AA pix^{-1}$. Final stacked on-source exposure
times were as follows: mask 1 - 19.8ksec, mask 2 - 19.1ksec, mask 4 -
10.8ksec, mask 730 - 32.4ksec, mask 444 - 45.5ksec, mask323A -
10.0ksec, mask323B - 7.2ksec. NOTES: (1) Identification numbers follow
Le Borgne, 1992. (2) Duplicate masks. These contained the same objects
except that the bright target \#323 on mask 323A was replaced on mask
323B by fainter targets to increase the number of spectra
obtained. (3) Numbers in parentheses denote exposures in poor
conditions.}
\label{tab-obslog}

\begin{tabular}{ccccccc}
\noalign{\smallskip}
\hline
\noalign{\smallskip}
\# Mask & Targets$^1$ & Dates & Total Exposure (ksec)$^3$  & Seeing(arcsec) \\
\noalign{\smallskip}
\hline
\noalign{\smallskip}
1    & 138,117,231,321,456,355,308,236,408			& 1995 May 27  & 10.8   	& 1.0 \\
     & ``							& 1995 May 29  & 9.0    	& 1.0--1.5 \\
2    & 129,171,229,328,384,468,492				& 1995 May 28  & 5.4    	& 1--2 \\
     & ``							& 1995 May 30  & 11.9   	& 1.0 \\ 
4 & 238,273,384,467						& 1995 June 1  & 10.8	& 0.7--1.0 \\
730  & 730,158,167,179,230,254,257,342,464,476,461 		& 1996 June 18 & 14.4   	& 1.1 \\
     & `` 					  		& 1996 June 19 & 7.2 (3.6) 	& 1.0 \\
     & ``							& 1996 June 20 & 7.2       	& 1.0 \\
444  & 444,117,145,236,242,256,262,297,306,307,308,344,381,431 	& 1996 June 16 & 7.5 (7.5) 	& 0.8--1.3 \\
     & ``							& 1996 June 17 & 15.0       	& 0.8--1.0\\
     & `` 							& 1996 June 21 & 18.0      	& 0.9--1.1\\
323A$^2$ & 323,132,159,167,190,200,205,223,288,295,317,321,389 	& 1996 June 15 & 10.0      	& 1.0--1.5\\
323B$^2$ & 277,132,159,167,190,200,205,223,288,295,257,321,389 	& 1996 June 20 & 7.2       	& 0.9\\
\noalign{\smallskip}
\hline
\noalign{\smallskip}
\end{tabular}
\end{table*}

The data were reduced using the LEXT package (Colless et al.\ 1990,
Allington-Smith et al.\ 1994). Frames were debiassed and median
combined taking into account small ($\sim0.3$ pixel $hr^{-1}$) shifts
due to instrument flexure. Dispersed tungsten flat fields were used to
correct for along-slit variations in illumination and pixel to pixel
sensitivity changes with wavelength. Third order polynomial fits to
CuAr arc lines provided wavelength calibration to a r.m.s. residual
of $\sim 2.0\pm0.1 $\AA. Sky subtraction was performed in two
steps. Firstly we fit the summed slit profiles, excluding the object
rows and edge effects.  The deviation from this overall profile was
then fit column by column to remove wavelength variations in the
profile. Finally the object rows were extracted with a Gaussian
weighting to optimize signal to noise (except for cases where the
target was partially merged with another object,
e.g. \#273). Redshifts were determined independently by two observers
(RSE and TMDE) using Karl Glazebrook's {\sc REDSHIFT} utility and are
summarized in Table \ref{tab-z}. We assigned a quality flag $q$ to
each spectrum according to the reliability of the redshift
determination. Redshifts determined from more than one strong line, or a
single line and continuum with good signal to noise (S/N) were assigned
$q=1$, designated `certain / probable'. Those determined from one or
more noisy lines were assigned $q=2$, designated `possible', and those
for which no redshift could be obtained were given $q=3$.

\begin{figure}
\psfig{file=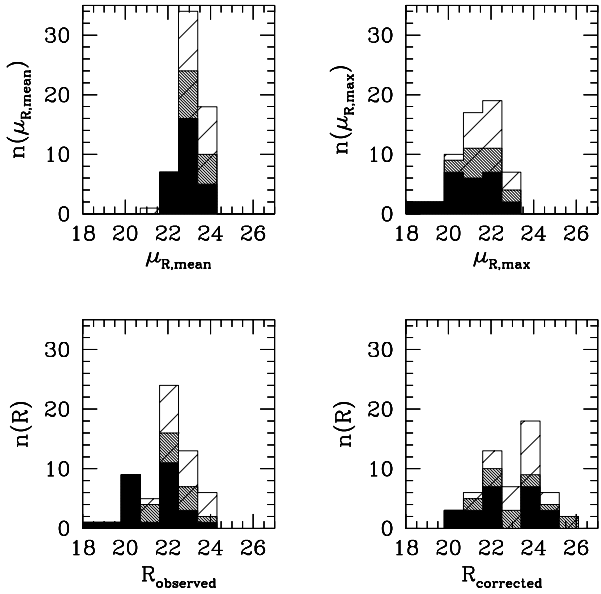,width=0.5\textwidth,angle=0}
\caption{Redshift completeness of the sample versus mean surface brightness
($\mu_{R,mean}$), peak surface brightness ($\mu_{R,max}$), observed
magnitude ($R_{observed}$) and magnitude corrected for lensing
amplification ($R_{corrected}$). Light, heavy and solid shading
represent arclets for which redshifts were undetermined, tentative and
firm respectively. It is evident that our success rate
for obtaining good redshifts falls as both the surface brightness and
total observed magnitude fall. However, the bottom right panel indicates
that, with the help of lensing amplification, we have obtained redshifts down
to $R\simeq25.5$.}
\label{fig-phothist}
\end{figure}

Of the 37 arclet candidates for which slits were cut, 7 failed to
yield spectra of adequate signal/noise (never reaching
$S/N_{cont}\simeq 3$). There are several reasons for this. Firstly, a
major limiting factor at these faint limits is the effectiveness of
sky subtraction. This is largely influenced by the surface brightness
of the target compared to that of the night sky (nominally
$\mu_{R,sky}\simeq 20.4\ mag\ arcsec^{-2}$ at La Palma in dark
time.)  Figure \ref{fig-phothist} shows the success rate of our survey
with both integrated $R$ magnitude and surface brightness,
$\mu_R$. It is evident that our sample is strictly limited at the
$\mu_R=24$ level and that our percentage success rate for determining
firm redshifts decreases as the mean surface brightness falls (100\%,
54\% \& 28\% in the three bins from $\mu_R=22$ to $\mu_R=24$
respectively.) 6 of the 7 low S/N arclets have surface brightnesses
$\mu_R \geq 23.2$ ($\leq 7\%$ of the night sky) and integrated
magnitudes of $B \geq 24.0$ ($R>22.0$), the faintest limit attained by previous
LDSS-2 surveys. In addition, in all but one case, less than 8
arcseconds of sky was available for sky subtraction due either to
contamination by a second galaxy in the slit or simply the length of
the slit. Finally, it is possible that misalignment of the slits from
their required positions could have led to low signal to noise in some
spectra. To investigate this, we took all the targets with $22<R<23$
and measured the average count level in each spectrum in a
$\sim1000$\AA\ wide region spanning the position of the R filter
($6500$\AA). If most slits were aligned well with the target objects,
there should be a correlation between the count level in the spectra
and R magnitude. We performed this test for 15 arclets and indeed found
a linear correlation of spectral counts with $10^{-0.4R}$ when three
outliers were discarded. At least one of the outliers could be
explained by poor slit orientation, suggesting that misalignment of
the slits with object positions was not a significant cause of light
loss in the majority of cases.

\begin{figure}
\psfig{file=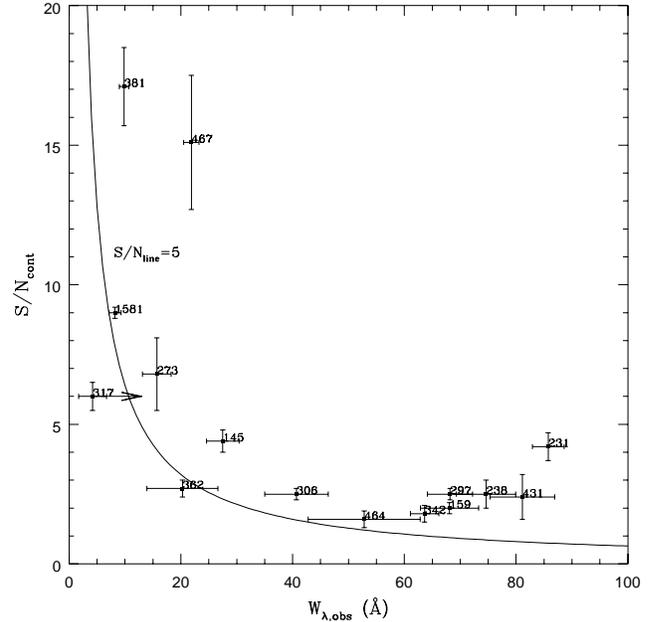,width=0.5\textwidth,angle=0}
\caption{Continuum signal to noise ratio versus observed [OII]
equivalent width for the arclets with $q=1$ (after Colless et al.,
1990). The curve shows a detection limit of $S/N_{line}=5$. Due to the
proximity of [OII] to the $\lambda5577$ sky emission line in \#317,
the equivalent width shown for this object is only a lower limit. Also shown
is one candidate with $q=2$ (\#362) which falls just below the
line. }
\label{fig-OIIsnw}
\end{figure}

Several spectra show good continuum signal to noise, but no
recognizable features and it is reasonable to ask what redshift ranges
are implied by the non-detection of features in these
spectra. Following Colless et al.\ (1990), we can use those spectra
where [OII] emission was detected to calculate our sensitivity to
detection of emission features. Given a continuum S/N value one may
obtain a limit on the minimum equivalent width of emission which could
result in a detection for a particular spectrum. As the continuum S/N
increases, we should become increasingly sensitive to lower equivalent
widths of emission. Figure \ref{fig-OIIsnw} shows a plot of the continuum
S/N adjacent to [OII] versus the observed equivalent width of [OII]
for the 15 arclets from which [OII] emission was detected. For an
unresolved line, continuum S/N ($S/N_{cont}$), line equivalent width
($W_{\lambda,obs}$) and line S/N ($S/N_{line}$) can be related to each
other, via simple assumptions, through the relation $S/N_{cont}
W_{\lambda,obs} \approx \sqrt{2\pi} \sigma_{line} S/N_{line}$. (We
assume the line profile is a Gaussian of sigma $\sigma_{line} \simeq
0.425 FWHM$. For our $\sim12$\AA\ resolution, this gives $\sigma_{line}
\simeq 5.1$.) Plotting this relation on the diagram shows that our
detection limit corresponds to a value of $S/N_{line} \simeq 5$. 

\begin{figure}
\psfig{file=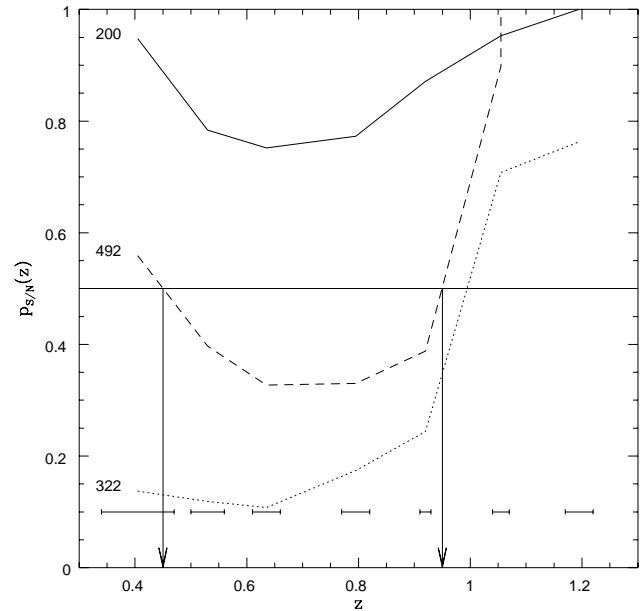,width=0.5\textwidth,angle=0}
\caption{Examples of the `$p(z)$' curves derived from the signal to
noise arguments given in \S2. The vertical arrows show how the
redshift constraint is defined. The three examples show one case
giving a single lower bound (\#322), one giving both upper and lower
limits (\#492) and one giving no bound at all (\#200). The horizontal
bars at the bottom of the plot show the size of the redshift windows
within which the probability value applies.} 
\label{fig-pz_eg}
\end{figure}

We now proceed to use this detection limit to put bounds on the
possible redshifts of those ($q=3$) arclets with good S/N but no
visible features. Given some signal to noise ratio for a piece of
continuum, one may calculate using the relation of Figure
\ref{fig-OIIsnw} the corresponding minimum equivalent width of
emission which could have been detected there. Combining this limit
with the distribution of [OII] equivalent widths seen in field
galaxies from other surveys, we may derive a probability that there
could be [OII] emission at this point without it having been
detected. The presence of [OII] at this point defines a redshift and
so this translates into a probability $p_{S/N}(z)$ that the galaxy
could have this particular redshift without any [OII] being
detected. Thus one may derive a $p_{S/N}(z)$ value for each point on
the spectrum once its $S/N_{cont}$ has been measured. At these faint
limits, the signal to noise in the red becomes prohibitive when one
encounters sky emission lines, and so we have chosen to apply this
measurement to 7 windows between sky emission, at: 5000-5550\AA,
5600-5800\AA, 6000-6200\AA, 6600-6800\AA, 7100-7200\AA, 7600-7700\AA,
\& 8100-8270\AA. These correspond to redshift windows for [OII] of
$z$=0.34-0.49,0.50-0.56,0.61-0.66,0.77-0.82,0.91-0.93,1.04-1.07 \&
1.17-1.22. Figure \ref{fig-pz_eg} shows 3 examples where $p_{S/N}(z)$
has been calculated.  To obtain a redshift constraint, we define a
threshold for $p_{S/N}(z)$. We can then use the point at which the
interpolated $p_{S/N}(z)$ curve rises above this threshold as a limit
on the redshift. Here we choose a value of $p_{S/N}(z)=0.5$ and quote
the redshift limits obtained for these $q=3$ spectra in Table
\ref{tab-z}.

There are several caveats to this process. Firstly, the value
$p_{S/N}(z)$ is only valid within the sky window where the signal to
noise was measured. It is possible that a significant equivalent width
of [OII] could remain undetected between these windows if the
continuum signal to noise in the sky band was low enough. Secondly, we
assume here that the distribution of observed equivalent widths
corresponds to those of the LDSS-2 deep redshift survey (Colless et
al. 1990). We note that this produces a conservative estimate since
our survey pushes significantly deeper in magnitude and redshift thus
making it likely that the true [OII] equivalent width distribution of
our sample contains more large equivalent widths than the one we have
used. This would lead to an over estimate of $p_{S/N}(z)$ for any
given continuum signal to noise. Finally, we point out that this
method will produce erroneous results if our sample does not
correspond to the general field in its distribution of equivalent
widths. To summarize, we have used our detection limit for emission
lines to derive redshift limits for those spectra which show good
continuum signal to noise but no features.

\begin{figure}
\psfig{file=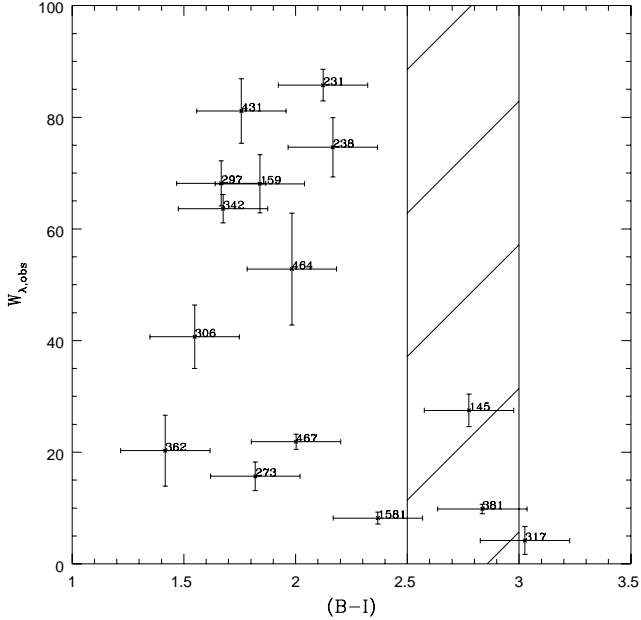,width=0.5\textwidth,angle=0}
\caption{The correlation between observed equivalent width and $(B-I)$
colour for those targets where [OII] was detected. The shaded area
encloses the region occupied by the E/S0 sequence of the
cluster. [OII] emission was detected in only one cluster member
(\#342) and in fact this object is well outside the cluster
sequence. It is thus clear that the blue colour selection criterion
proved a good predictor of the presence of [OII] .}
\label{fig-OIIBI}
\end{figure}

\begin{figure}
\psfig{file=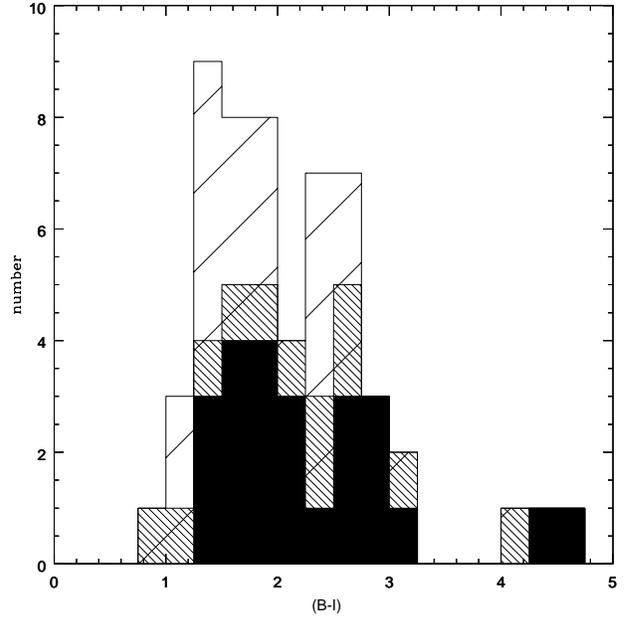,width=0.5\textwidth,angle=0}
\caption{The $(B-I)$ distribution of the spectroscopic sample. As
before, light, heavy and solid shading represent undetermined,
tentative and confirmed redshifts respectively. } 
\label{fig-BIhist}
\end{figure}

Figure \ref{fig-BIhist} shows how our redshift completeness varies
with the target colours. It can be readily seen that more redshifts
were obtained for blue objects than red. This probably reflects the
greater strength of [OII] emission in the blue objects as hypothesised
in our selection method. This is confirmed in Figure \ref{fig-OIIBI}
which shows the correlation between equivalent width in [OII] and
$(B-I)$ colour. Here, the shaded region marks the cluster sequence
colour. [OII] emission was detected for only one cluster galaxy which,
in fact, lies well outside the cluster sequence. There is a clear
anticorrelation between $(B-I)$ colour and equivalent width such that
the reddest galaxies have little or no [OII] emission. This confirms
the results seen by Colless et al.\ (1990) and Glazebrook et
al. (1995). We conclude that our colour selection was a useful way to
increase the likelihood of detecting [OII] emission (and thus
obtaining a redshift), as well as a method for selecting non-cluster
objects.

Of the 37 arclet targets, we were able to obtain a total of 32
redshifts, 18 of which were of good quality ($q=1$) and were behind
the cluster ($z_s>0.2$). Figures \ref{fig-spec1}, \ref{fig-spec2},
\ref{fig-spec3} and \ref{fig-spec4}, show the extracted spectra,
2-dimensional spectra and zooms of the arclets, while the redshift
catalogue is presented in Table \ref{tab-z}.

Several of the spectra merit individual discussion especially with
regard to their redshift determinations. First we deal with the
multiple images and follow with those that are singly imaged. 

\begin{description}

\item[{\bf 289} (R=20.5, z=1.034) \& {\bf 359}, (R=20.3, z=0.702)] The
spectra for these giant arcs are discussed by Pell\'o et al.\ (1992).

\item[{\bf 328} (R=22.0, z=0.702) \& {\bf 389} (R=21.5, z=0.702)] Both
part of the \#359 multiple system showing weak features corresponding to
the \#359 redshift of $z=0.702$. \#389 appears seen through the disc of
the spiral galaxy \#373 and consequently the features are so weak that
we assign $q=2$.

\item[{\bf 273} (R=21.9, z=0.800)] The spectrum of this arc is heavily
contaminated with light from the nearby elliptical \#268 (see Figure
\ref{fig-spec2}). However, as may be seen from the 2-d spectrum, the
emission line is very prominent at the lower edge of the \#268
continuum. Only the 3 rows containing the emission line were extracted
and no Gaussian weighting was used.

\item[{\bf 384} (R=21.2, z=2.515)] The spectrum of this spectacular
arc has already been discussed in a recent letter (Ebbels et
al. 1996.) Many rest frame ultra-violet metal lines are seen, along
with Lyman-$\alpha$ in absorption at $z=2.515$.

\item[{\bf 444} (R=22.7, z=1.030)] The spectrum has good continuum
signal to noise and is extremely blue. We base our identification on
the FeII doublet at $\lambda2587$ and $\lambda2600$. This is in good
agreement with the predicted redshift of $z=1.1\pm0.1$. However this
remains a tentative ($q=2$) identification.

\item[{\bf 145} (R=22.1, z=0.628)] [OII] is seen strongly in this
spectrum, along with some other more tentative features.

\item[{\bf 158} (z=0.721)] This spectrum could be a blend of 2 sources
as may be inferred from Figure \ref{fig-spec1}. We identify several
features (principally [OII]) at a redshift of $z=0.723$ while other
features, including the continuum, are more consistent with a redshift
of $z=0.167$. We suggest that most of the continuum comes from a
bright cluster member while the fainter (by $0.7mag$), diffuse object
is a background galaxy (B=22.5, henceforth denoted \#1581) and
produces the [OII] emission. Unfortunately the ground based seeing
precludes any spatial identification of the emission line with one
particular component.

\item[{\bf 159} (R=21.9, z=0.564)] [OII] emission dominates this
spectrum which also includes several absorption features.

\item[{\bf 205} (R=20.3, z=0.693)] This spectrum has good signal to
noise with both absorption and emission features around the Balmer
break visible.

\item[{\bf 231} (R=22.0, z=0.563)] The ground-based seeing causes this
spectrum to be a blend of 231 and compact component. These have R
magnitudes of 22.0 and 23.4 respectively, so that no more than 21\% of
the light can be contributed by the compact component and it is
unlikely that the emission line arises from it alone.

\item[{\bf 238} (R=21.9, z=0.635)] A single emission feature
identified with [OII] dominates this spectrum.

\item[{\bf 242} (R=21.8, z=654)] Again, [OII] emission is seen along
with several absorption features at lower signal to noise.

\item[{\bf 297} (R=22.4, z=0.450)] [OII] emission is very strong here
and is accompanied by tentative identifications of the
$H\beta$/[OIII] triplet.

\item[{\bf 306} (R=23.7, z=0.450)] [OII] emission along with several other more
tentative identifications lead to the redshift for this arclet.

\item[{\bf 317} (R=20.1, z=0.474)] The [OII] emission here falls very
close to the $\lambda5577$ sky line and so we can determine a lower
limit only on its equivalent width. However, several other features
confirm the redshift.

\item[{\bf 381} (R=20.5, z=0.521)] This spectrum, which has one of the
best signal to noise ratios of all those obtained, exhibits many
features from MgII absorption to $H\beta$.

\item[{\bf 431} (B=23.5, z=0.675)] Despite its proximity to the
$\lambda6300$ night sky emission line, the [OII] emission in this
arclet is strong and unmistakable on the 2-d spectrum.

\item[{\bf 456} (R=22.2, z=0.538)] Several absorption features
including the K/H doublet are visible with good enough signal to noise
to obtain a redshift for this arclet.

\item[{\bf 464} (R=22.9, z=0.476)] A strong emission feature
identified with [OII] plus a possible G absorption band lead to a
redshift for this arclet.

\item[{\bf 467} (R=20.5, z=0.475)] Excellent signal to noise provides
a sure redshift for this arclet, with up to 14 identifications from
[OII] emission to $\lambda5268$ absorption.

\end{description}

We also mention the object \#323 which, being very elongated but at an
angle $\gtapprox 45^{\circ}$ from the shear direction, would cause
problems for the model were it at a redshift much higher than the
cluster. Identification of several absorption features in this
spectrum reveal it to be a cluster member at a redshift of $z=0.179$,
thus posing no threat to the model. Finally, we note that of our new
redshifts, 7 arclets are found in three redshift structures. The
sensitivity to small-scale redshift clustering is an artifact of the
small angular extent of the survey and is a factor affecting all
small-area surveys (including the photometric redshift surveys in the
Hubble Deep Field, see \S5). Thus, gaining a statistically-reliable
view of the field redshift distribution will require inversion through a
number of lenses.

The 18 new redshifts bring the total number of arclets with good
redshifts in Abell 2218 to 20 -- making it the largest number for any
lensing cluster to date. Table \ref{tab-z} lists the results in order
quality with multiple image systems appearing in the top 2
sections. The following analysis of the inversion method and our
conclusions are based solely on the best quality ($q=1$) arclets,
though we expand the sample to include the $q=2$ arclets where
appropriate. For completeness we also list the $q=3$ arclets and their
redshift limits from the signal to noise argument above in order to give
an indication of the sample which might be reached with a 10m class
telescope.

\begin{figure*}
\caption{LDSS-2 spectra for arclets in Abell 2218, listed in order of
identification number. All spectra with $q=1$ are shown, while arclets
\#362, \#389, \& \#730 are included for interest. (Arc \#730 is shown
with a rest frame wavelength scale corresponding to its predicted
redshift.)  See text for individual discussion. The spectra have been
fluxed and smoothed to the effective resolution of LDSS-2, but are not
corrected for atmospheric absorption. }
\label{fig-spec1}
\end{figure*}
\begin{figure*}
\caption{LDSS-2 spectra for arclets in Abell 2218. (Continued)} 
\label{fig-spec2}
\end{figure*}
\begin{figure*}
\caption{LDSS-2 spectra for arclets in Abell 2218. (Continued)} 
\label{fig-spec3}
\end{figure*}
\begin{figure*}
\caption{LDSS-2 spectra for arclets in Abell 2218. (Continued)} 
\label{fig-spec4}
\end{figure*}

\begin{table*}
\centering

\caption{Arclet Spectroscopy in Abell 2218. We list first the multiple
image systems, followed by the singly imaged ones. Only arclets with
$z_s\geq0.2$ are listed, except in the case of \#323 (see text). The
arclets for which no redshift prediction was possible were added to
the sample at points on the masks where suitable bright arclets with
inverted redshifts were unavailable. 
COLUMN NOTES: (1) $R_{corr}$ calculated assuming $z=z_{opt}$. (2)
Quality (q): 1=certain/probable, 2=possible, 3=limit only (3) From
WFPC-2 F702W image transformed to R as detailed in KESCS. Estimated
error $0.2$ mag. (4) For those objects with undetermined redshifts, we
quote limits on the redshift derived from the argument in \S2. 
COMMENTS: (1) Class III; No maximum in $p(z)$:
$z_{opt}\rightarrow\infty$ (2) Class I; No maximum in $p(z)$:
$z_{opt}\leq0.2$ (3) Serendipitous object (4) Arclet off WF frame,
photometry quoted in B band (5) Redshifts determined by Pell\'o et
al., 1992. }
\label{tab-z}
\begin{tabular}{cccccccccclc}

\noalign{\smallskip}
\hline
\noalign{\smallskip}

\# & $\mu_R$ & R$^{(3)}$ & $R_{corr}^{(1)}$ & $z_{spec}^{(4)}$  & $z_-$ & $z_{opt}$ &
$z_+$ & $\delta_z/z_{opt}$ & quality (q)$^{(2)}$ & identified features & comment \\

\noalign{\smallskip}
\hline
\noalign{\smallskip}

289 & 23.1 & 20.5 & 22.3 & 1.034 &     & 1.034 &      &       & 1 & $[OII]$ & 5\\
359 & 22.7 & 20.3 & 24.9 & 0.702 &     & 0.702 &      &       & 1 & $K,H,H\gamma$ & 5\\

\hline

328 & 22.6 & 22.0 & 24.9 & 0.702 &     & 0.702 &      &  0.00 & 1 & $K?,H,4000$ & \\
389 & 22.8 & 21.5 & 24.9 & 0.702 &     & 0.702 &      &  0.00 & 2 & 
$K?,H?$ & \\
384 & 23.4 & 21.2 & 24.1 & 2.515 & 2.6 & 2.8   & 3.3  & -0.10 & 1 &
$Ly-\alpha,SiII,OI,CII,SIV,CIV$  & \\			      
444 & 23.9 & 22.7 & 25.7 & 1.030 & 1.0 & 1.1   & 1.2  & -0.06 & 2 & $FeII?$ &
\\

\hline

145 & 23.3 & 22.1 & 22.3 & 0.628 & 0.5 & 0.9   & 2.9  & -0.3  &
1 & $[OII],K?,H?,H\delta?$ & \\						      
1581&      & 22.5 & 23.0 & 0.721 & 0.3 & 0.6   & 1.4  &  0.20 &
1 & $MgII?,[OII]$ & 4\\						      
159 & 22.7 & 21.9 & 21.9 & 0.564 & 0.2 & 0.2   & 0.2  &  -   &
1 & $[OII],H\theta?,H\eta?,H\zeta?,H?,[OIII]?$ & 2 \\						      
205 & 22.4 & 20.3 & 21.1 & 0.693 & 0.2 & 0.3   & 0.4  &  1.3  &
1 & $[OII],H\theta,H\eta,H\zeta,K,H,H\delta$ & \\						      
231 & 23.1 & 22.0 & 22.2 & 0.563 & 0.3 & 0.4   & 0.6  &  0.4  &
1 & $[OII],K?,H?,4000$ &\\				      
238 & 23.5 & 21.9 & 23.9 & 0.635 & 0.8 & 1.2   & 1.6  & -0.5  &
1 & $[OII],H?,4000?$ & \\					      
242 & 23.2 & 21.8 & 21.8 & 0.654 & 0.2 & 0.2   & 0.2  &   -   &
1 & $[OII],H\zeta?,K?,H?,H\delta?$ & 2\\			
273 & 22.2 & 21.9 & 23.8 & 0.800 & 0.5 & 0.6   & 0.7  &  0.33 & 1 & $[OII]$ &\\
297 & 23.7 & 22.4 & 23.8 & 0.450 & 0.5 & 0.6   & 0.7  & -0.3  &
1 & $[OII],H\delta?,H\beta,[OIII]$ & \\						      
306 & 22.8 & 23.7 & 24.2 & 0.450 &  -  &  -    &  -   &   -   &
1 & $[OII],H\zeta,H?,G?,H\beta,[OIII]$ & 1 \\						      
317 & 22.8 & 20.1 & 20.6 & 0.474 & 0.2 & 0.3   & 0.4  &  0.6  &
1 & $[OII],H\eta,H\zeta,K,H,4000,H\delta$ & \\						      
381 & 22.7 & 20.5 & 21.0 & 0.521 & 0.3 & 0.4   & 0.5  &  0.3  &
1 & $MgII?,[OII],H\theta,H\eta,H\zeta,H\delta,G$ & \\						      
431 &      & 23.5 &      & 0.675 &  -  &  -    &  -   &   -   &
1 & $MgII?,MgI?,[OII]$ & 4\\						      
456 & 23.4 & 22.2 & 24.5 & 0.538 & 0.4 & 0.6   & 0.8  & -0.1  &
1 & $[OII]?,H\zeta?,K,H,4000,G?$ & \\			      
464 & 23.3 & 22.9 & 24.8 & 0.476 & 0.9 & 1.1   & 1.3  & -0.6  &
1 & $[OII],G?$ & \\						      
467 & 22.1 & 20.5 & 21.7 & 0.475 & 0.4 & 0.4   & 0.5  &  0.2  &
1 & $[OII],H\theta,H\eta,H\zeta,K,H,4000,H\delta$, \\
    &      &      &      &       &     &       &      &       & & $G?,H\beta?,[OIII]?,Mgb?,5268?$\\
132 &    & 23.7 & 23.7  & 0.703 & 0.2 & 0.2   & 0.2  &   -   & 2 &
$MgI?,MgII?$ & 2,4\\
190 & 23.4 & 22.1 & 22.1 & 0.708 & 0.2 & 0.2   & 0.2  &   -   & 2 &
$[OII]?$ & 2\\
229 & 23.8 & 23.8 & 24.8 & 0.830 & 0.8 & 1.0   & 1.2  & -0.2  & 2 &
$MgI?,MgII?$ & \\
236 & 23.3 & 22.4 & 23.1 & 0.570 & 0.4 & 0.5   & 0.6  &  0.1  & 2 &
$H\theta?,H\eta,K?,H\gamma?$ & \\
262 & 23.7 & 21.8 & 21.8 & 0.596 & 0.2 & 0.2   & 0.2  &  -    & 2 &
$K?,H?,4000?,H\delta?,H\beta?$ & 2\\
307 & 23.2 & 22.1 & 23.1 & 0.390 & 0.3 & 0.3   & 0.4  &  0.3  & 2 &
$K?,4000?,G?,Mgb?$ & \\
355 & 23.7 & 23.0 & 23.5 & 0.470 & 0.3 & 0.4   & 0.6  &  0.2  & 2 & $G?$& \\
362 & 23.9 & 22.6 & 25.8 & 0.532 & 0.5 & 1.1   & 2.0  & -0.5  & 2 & $[OII]?$ &\\
    &      &      &      &       &     &       &      &       &   & & \\
117 & 23.1 & 23.0 & 23.4 &   -   &  -  &  -    &  -   &       & 3 & & 1\\
129 & 22.8 & 23.4 & 23.8 &   -   & 0.3 & 0.5   & 0.8  &       & 3 & & \\
167 & 23.7 & 22.1 & 22.4 & $>$0.44 & - &  -    &  -   &       & 3 & & 1\\
171 & 23.3 & 22.3 & 23.0 &   -   & 0.4 & 0.6   & 0.8  &       & 3 & & \\
179 & 23.4 & 21.9 & 22.1 &   -   &  -  &  -    &  -   &       & 3 & & 1\\
200 & 23.2 & 23.5 & 25.0 &   -   & 0.9 & 1.1   & 1.4  &       & 3 & & \\
223 & 22.9 & 23.1 & 24.1 & $<$0.53,$>$0.72 & 0.6 & 0.7   & 0.9  &       & 3 & & \\
230 & 23.3 & 22.6 & 24.0 &   -   & 0.4 & 0.5   & 0.6  &       & 3 & & \\
254 & 23.9 & 22.0 & 23.9 &   -   & 0.4 & 0.5   & 0.7  &       & 3 & & \\
256 & 23.2 & 23.9 & 24.1 &   -   & 0.2  & 0.4  & 0.8  &       & 3 & & \\
257 & 23.2 & 21.9 & 22.7 & $>$0.90 & 0.3 & 0.5   & 0.8  &       & 3 & & \\
277 & 23.3 & 22.4 & 23.4 & $>$0.95 & 0.3 & 0.4   & 0.5  &       & 3 & & \\
308 & 23.4 & 22.2 & 23.4 & $>$0.95 & 0.5 & 0.6   & 0.8  &       & 3 & & \\
322 & 21.3 & 21.1 & 21.1 & $>$0.99 & 0.2  & 0.2  & 0.2  &       & 3 & & 2,3\\
344 & 23.8 & 22.9 & 22.9 & $>$0.92 & 0.2 & 0.2   & 0.2  &       & 3 & & 2\\
390 & 23.7 & 24.1 & 24.1 &   -   & 0.2  &  0.2  & 0.2   &       & 3 & & 2,3\\
468 & 23.5 & 22.6 & 23.6 & $>$0.81 & 2.6 & 2.8   & 3.3  &       & 3 & & \\
476 & 23.7 & 23.1 & 23.8 & $<$0.45,$>$0.75 & 0.3 & 0.3   & 0.4  &       & 3 & & \\
492 & 22.8 & 21.6 & 21.7 & $<$0.45,$>$0.95 & 0.2 & 0.2   & 0.2  &       & 3 & & 2\\
730 & 23.7 & 22.8 & 25.1 & $>$0.57 & 1.0 & 1.1   & 1.2  &       & 3 & & \\
    &      &      &      &       &     &       &      &       &   & & \\
323 & 22.7 & 20.1 & 20.1 & 0.179 & 0.2 & 0.2   & 0.2  & -0.1  & 2 &
$K,4000,Mgb,NaD,H\alpha$ & \\
\noalign{\smallskip}						
\hline								
\noalign{\smallskip}						
\end{tabular}
\end{table*}

\section{Comparison with Lensing Predictions}

We now compare the redshifts obtained with those predicted by the lens
inversion method outlined in KESCS. 

The arcs whose redshifts can be predicted fall into two classes:
multiple images and single images. The redshift predictions for the
multiple images have much smaller uncertainties than those for
singly-imaged sources because of the larger number of constraints
implied when several images of the same source are taken together. It
is therefore particularly important to prove that that the inversion
works for these sources. However, in the case of a singly-imaged
arclet, the uncertainty in the redshift prediction is more typical of
that expected for the larger fainter population of arclets addressed
by KESCS. The uncertainty here depends on the redshift, the size of
the local shear and the unknown intrinsic shape of the source. The
average uncertainties listed in KESCS were $\delta z /z = 22\%$ for
multiple images and $\delta z /z = 30\%$ for single images.

In order to understand what level of agreement we expect between our
inverted redshifts and measured spectroscopic ones, it is useful to
consider the method by which the inverted redshifts are obtained. For
any particular singly imaged arclet at any point on our image, our
mass model can predict magnitude and direction of the shear induced by
the lens. The magnitude of this shear depends on the redshift of the
arclet and increases from a minimum at the cluster redshift to an
absolute maximum as the arclet redshift approaches infinity (This
limit is actually replaced by a redshift of $\sim4$ in our model). The
observed shape of the arclet is a combination of its intrinsic shape
and this redshift dependent distortion. Conversely, knowing the
observed shape, and given a redshift, one can predict the intrinsic
shape using the calculated shear. As we move the arclet in redshift
from the distance of the cluster out to infinity, its predicted
intrinsic shape will change, eventually tending to a constant shape at
large redshift. 

\begin{figure*}
\psfig{file=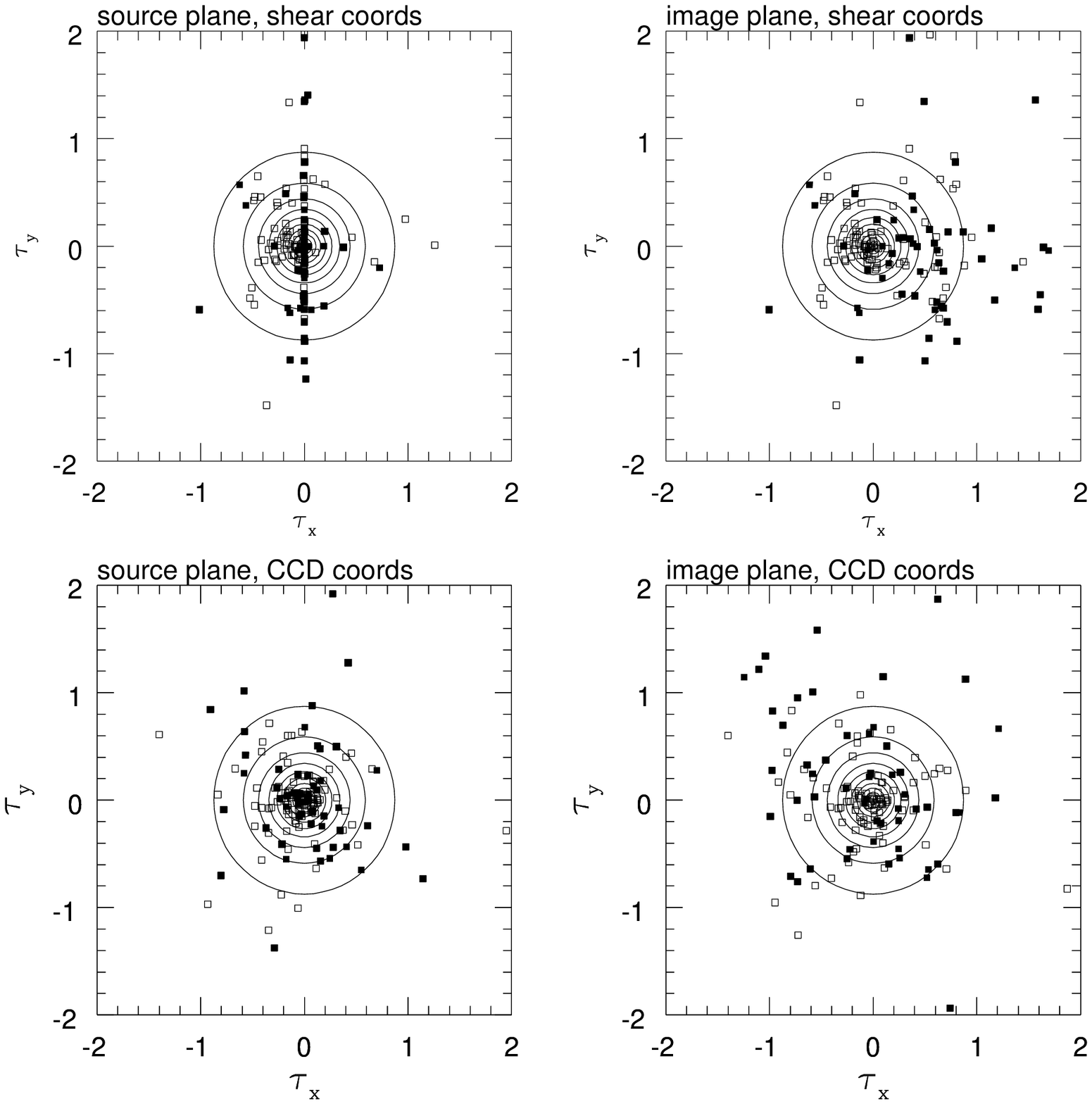,width=1.0\textwidth,angle=0}

\caption[]{The lens inversion method. We show plots of the $\bar{\tau}$
ellipticity vectors for the arclets in the source plane ({\it left
panel}) and the image plane ({\it right panel}). For each plane, we
show the $\bar{\tau}$ vectors with respect to the frame of the local
shear ({\it upper panel}) and the CCD frame ({\it lower panel}). In
all plots we show linearly spaced contours of the field galaxy
$\bar{\tau}$ distribution as measured from HST field survey images
(Ebbels et al. in preparation). Solid symbols represent the
spectroscopic sample while open ones indicate the full sample of
objects with areas over 50 pixels (137 arclets.) The inversion method
is clearly seen at work in the upper panel. In the image plane ({\it
top right}), most of the arclets have $\tau_x>0$ since most of them
(especially the spectroscopic sample) are aligned with the local
shear. Placing each source at a redshift behind the cluster translates
the corresponding point on this diagram in x only. For each arclet we
choose that redshift at which $\tau$ is minimized, and thus the
invertible arclets (class II) will end up on the line $\tau_x=0$ in
the source plane ({\it top left}). In the source plane, class I
arclets (not sufficiently distorted to invert) lie in the region
$\tau_x<0$ and class III arclets (too elongated to invert) lie in the
region $\tau_x>0$ as is clearly seen. Moving to the CCD frame, one can
see that the distribution of points in the image plane ({\it bottom
right}) is more extended than that of field galaxies in general
(justifying their selection as elongated arclets.) In the source plane
({\it bottom left}) the effect is reversed since we choose the
roundest shape for each object and the distribution of points is more
peaked than the field. It is clear that when referred to the CCD
coordinates, there is no preferential direction, even in the source plane.
}

\label{fig-plottau}
\end{figure*}

We may represent the shape of any object on a plot of ellipticity
vectors $\bar{\tau}$ as shown in Figure \ref{fig-plottau}. Using the
notation of KESCS, $\bar{\tau} = \tau e^{2i\theta}$ and $\tau = (a^2 -
b^2)/(2ab)$ with $a$ and $b$ the semi-major axes of the equivalent
ellipse. The vectors may be plotted with respect to the axes of the
CCD (large plots) or with respect to the axis of local shear for each
arclet (insets). In these local shear axes, the lensing transformation
conserves the y component of the ellipticity, $\tau_y$ and thus the
process of moving the arclet in redshift simply corresponds to moving
its point in the $-x$ direction across the $\bar{\tau}$-diagram. As we
move the arclet out in redshift and thus to the left in $\tau_x$
across the plot, we can ask what is the likelihood that any given
galaxy in the field would have the shape defined by this point on the
$\bar{\tau}$ diagram. This likelihood will correspond to the value of
the ellipticity distribution of field galaxies, $p(\bar{\tau})$
(represented by the contours on the plots) at that point. We can
therefore define a redshift probability function $p(z)$ for each
arclet, corresponding to the likelihood that its intrinsic shape for
any redshift would be found in the general population of field
galaxies. The $p(\bar{\tau})$ distribution peaks at an ellipticity of
$\tau=0$ and has circular symmetry. Thus the maximum in our
probability, $p(\tau_x)$ or correspondingly, $p(z)$ will occur when
$\tau_x=0$. We therefore assign a most likely redshift $z_{opt}$ to
each arclet defined by the redshift which places it at $\tau_x=0$ on
the diagram of Figure \ref{fig-plottau}.

This method divides the singly imaged arclet population into three
classes, two of which preclude a determination of $z_{opt}$ (see also
Figure 7 of KESCS). The first, class I, corresponds to those arclets
with $\tau_x < 0$. These cannot be moved to $\tau_x=0$ by placing them
behind the cluster and must therefore be either cluster or foreground
galaxies, or background galaxies which are not distorted enough to lie
within a $45^{\circ}$ cone about the axis of local shear. Therefore no
meaningful $z_{opt}$ may be determined for them. The second, class II
arclets constitute those which may be moved to $\tau_x=0$ by placing
them at a particular redshift behind the cluster and a determination
of $z_{opt}$ is possible for these arclets. Finally, particularly in
areas away from the cluster core where the shear is weak, we have some
arclets (class III) which although aligned closely with the shear
axis, do not reach $\tau_x=0$ even at infinite redshift. Again, no
meaningful $z_{opt}$ can be given for these arclets. 

We must emphasize that the redshift prediction $z_{opt}$ only has
physical meaning in a statistical sense. That is, while the shape
corresponding to $z_{opt}$ may not in fact be the {\it actual} intrinsic
shape of the object, the {\it distribution} of these shapes across the
population of inverted objects must be drawn from the distribution of
field galaxy shapes, $p(\bar{\tau})$. If this is true, then while the
individual $z_{opt}$ values may not always correspond to the true
redshift, the redshift properties of the population {\it as a whole} (such
as $<\! z\! >$ and shape of the $n(z)$) should be correct. Later we assess
the manner in which our inverted shape distribution (Figure
\ref{fig-plottau}, {\it left}) samples the field galaxy $p(\bar{\tau})$ but
first we discuss individual cases of inversion, starting with multiple
images.

In KESCS' model, seven multiply imaged systems were used to constrain
the mass model. These were identified using similarities in their HST
morphologies, colours, surface brightnesses and on general assumptions
about the lensing geometry. Only two of these (\#359 and 289) already
had redshifts from the spectroscopic study of Pell\'o et al.\ (1992).
Furthermore, none had actually been confirmed spectroscopically as
multiple. Thus our redshift for \#328, a component of the 5-image
\#359 system is a strong vindication of the mass model and the ability
of HST to identify multiple images. The spectrum of \#389 is
consistent with its being a third image of is system although highly
contaminated by light from the obscuring cluster spiral \#373. Indeed,
the identification of this as a 5-image system was the major
motivation of KESCS to extend their model to include galaxy scale mass
components. These 5 images could not be modeled using cluster scale
mass components alone. Therefore, this confirmation reinforces our
confidence in the mass model down to very small scales
($\sim75kpc$). The \#384 system consists of one merging image pair
(arc \#384) and its counter-image (arc \#468). Ebbels et al.\ (1996)
confirmed that both halves of the merging pair have the same redshift
($z=2.515$) and also found the spectrum of \#468 to be compatible with
this redshift, although its signal to noise is poor. Of particular
significance, and also discussed briefly in Ebbels et al., is the fact
that the spectroscopic redshift of this system is in excellent
agreement with the lensing prediction: $z=2.8^{+0.5}_{-0.3}$
(KESCS). The redshift of this system is particularly important due to
its proximity to a critical line, meaning that the predicted redshift
is extremely sensitive to changes in the mass model. In summary,
therefore, four images from two multiple systems are consistent with
the earlier mass model and the most highly-magnified has a redshift
within the expected uncertainties.

\begin{figure}
\psfig{file=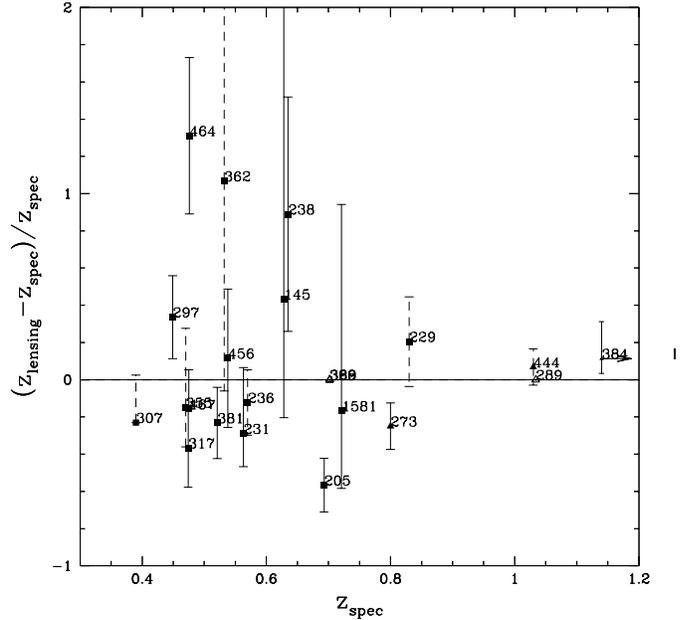,width=0.5\textwidth,angle=0}
\caption{The fractional difference between $z_{lensing}$ and
$z_{spec}$ for the A2218 sample. Single images are represented by
squares and multiple images by triangles. The open triangles show
multiple images that were used as constraints on the model. Objects
with $q=1$ are shown with solid error bars, while those with $q=2$
have dashed error bars. }
\label{fig-dzz}
\end{figure}

Turning to the singly-imaged arclets, Figure \ref{fig-dzz} shows the
fractional difference between $z_{opt}$ and $z_{spec}$ for the 
arcs and arclets where both a spectroscopic and inversion redshift above
$z=0.2$ are available. Here the error bars refer to the 80\%
likelihood values discussed by KESCS (henceforth denoted by $\sigma_{80\%}$)
and we include arclets with $q=2$ for purposes of illustration.
Although the sample of $q=1$ arclets is small, it is useful to ask at
what level of significance do the results agree with the predictions.

The arclets can be grouped according to the degree of success with which we
predicted their redshifts. In total there are 11 arclets of class II
and $q=1$ for which meaningful redshift predictions can be made. Of these, 9
match their spectroscopic redshifts within $2\sigma_{80\%}$. As tests
of the method, the best of these are arclets \#231, \#297, \#317,
\#381 \#467 \& \#456 which have small error bars leaving little
likelihood that the predictive success could be a product of chance
alone. The remaining 3 arclets (\#145, \#1581 \& \#238) have larger
error bars due to their relatively large values of $|\tau_{y}|$, but
still agree tolerably well with their spectroscopic redshifts. Finally
there are two arclets whose predictions do not match their
spectroscopic redshifts within $2\sigma_{80\%}$ - objects \#205 and
\#464. Morphologically, \#205 is clearly a spiral galaxy (Figure
\ref{fig-spec2}) and this apparent lack of distortion so close to the cluster
core leads to the low redshift prediction. \#464, however appears
quite distorted in the direction of the local shear and its predicted
redshift is correspondingly high. The disagreement between prediction
and measurement for the arclets \#205 and \#464 must therefore result
from the natural dispersion of image shapes on the sky, leading to
chance alignments in the source plane away from and along the shear
direction respectively. Given the shape of the field galaxy
ellipticity distribution as illustrated by contours in Figure
\ref{fig-plottau}, it is not unusual to find galaxies in these regions
(reasonably large $|\tau_{x}|$ but small $|\tau_{y}|$). Thus we
conclude that these two cases represent merely a statistical
fluctuation in this small sample.

We can now consider, in the light of Figure \ref{fig-dzz}, the nature of
the error in the predictions for the population as a whole. Taking
only the $q=1$ arclets, the distribution shown in Figure \ref{fig-dzz}
has a mean of $<\! \delta_z/z\! > = 0.07$ and dispersion
$\sigma_{\delta_z/z} = 0.48$, while including the $q=2$ arclets brings
these values to $<\! \delta_z/z\! >=0.09$ and $\sigma_{\delta_z/z} =
0.41$. We therefore expect the fractional error in the predicted
quantities of the whole population (such as $<\! z\! >$) to be of order 10\%
or less, despite the dispersion of individual points.

The conclusion from the above discussion is that while $z_{opt}$ and
$z_{spec}$ agree well for many of the arclets, we cannot expect this
to be the case for all arclets. This is simply due to the unknown
position of $\bar{\tau}$ within the natural distribution of source
shapes.  As mentioned above, it is useful to consider how the
predicted ellipticity distribution of our sample compares with that
seen in the field generally. Returning to Figure \ref{fig-plottau},
the lower two panels illustrate how both the spectroscopic and fainter
samples trace the field galaxy distribution (contours, measured from
$\sim10000$ galaxies between $I=18-25$ taken from HST WFPC2 fields in
the MDS and Groth Strip Surveys (Ebbels et al. in prep.)). Two effects
can clearly be seen. The first is the tendency in the image plane, for
the spectroscopic sample (filled symbols) to be more extended than the
contours. This justifies our selection of these objects as arclet
candidates -- they are more extended than the field population in
general. The second effect is a result of the method itself,
specifically that our choice of the optimum redshift of an object will
correspond to its lowest obtainable ellipticity. Thus the distribution
of points in the source plane will tend to be more centrally
concentrated than in the field and this can readily be seen in the
Figure. Notwithstanding these biases however, it is clear that our
predicted distribution favors no particular orientation and is
centrally peaked -- the two features most noticeable about the unlensed
field distribution.

\begin{figure}
\psfig{file=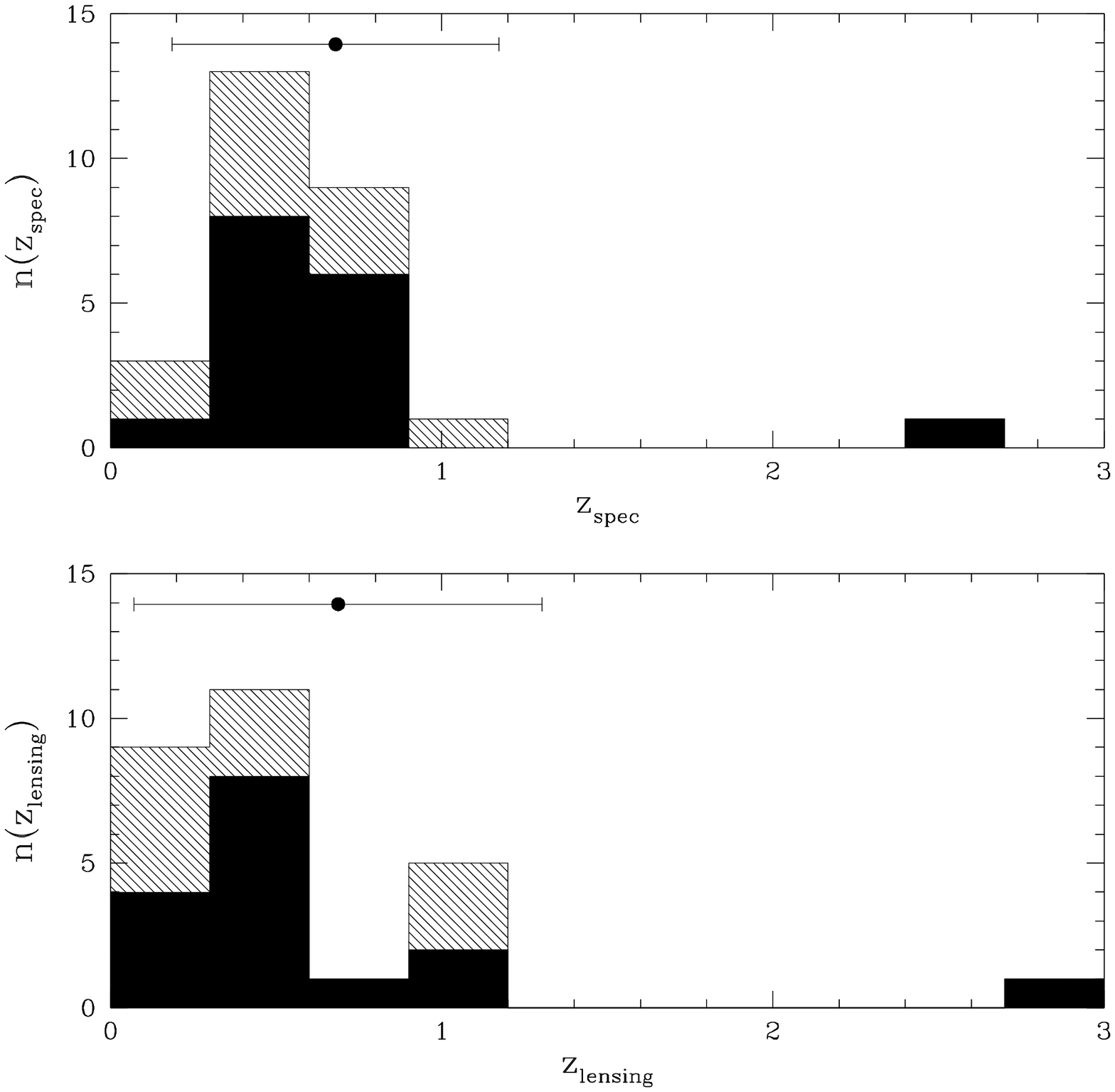,width=0.5\textwidth,angle=0}
\caption{The distribution of our sample with spectroscopic ({\it
left}) and inversion ({\it right}) redshifts. Objects with
firm ($q=1$) and tentative ($q=2$) redshifts are denoted
by solid and heavy shading shading respectively. Here we have ensured
that both histograms correspond to the same sample of objects which include
multiple images (except those used as model constraints). In each plot
the point with error bar marks the mean and dispersion of the
distribution for the $q=1$ sample.
} 
\label{fig-zhist}
\end{figure}

In using this technique to probe the redshift distribution of faint
galaxies, we are mainly concerned with certain statistical properties
of this distribution such as the mean $<\! z\! >$ and the dispersion
$\sigma_z$. We are therefore compelled to compare the predicted values
of these quantities with their actual values for our sample. We show
the redshift distributions derived from the inversion method and from
spectroscopy in Figure \ref{fig-zhist}, where we have ensured that
both distributions correspond to that subset of our sample possessing
both inversion and spectroscopic redshifts. Clearly the two
distributions are similar in form with the bulk of the arclets below a
redshift of one. Using just those arclets with $q=1$, we find $<\!
z_{spec}\! > = 0.68$ while the predicted value is $<\! z_{opt}\! > =
0.69$, a difference of just 1.5\%. Expanding the sample to include the
$q=2$ arclets, we find $<\! z_{spec}\! > = 0.63$ and $<\! z_{opt}\! >
= 0.61$, a difference of 2\%. In this case therefore, the mean
redshift is very accurately predicted by the technique. Looking at the
shape of the distributions, one may compare their
dispersions. Considering the $q=1$ data, we find
$\sigma_{z_{spec}}=0.49$ and $\sigma_{z_{opt}}=0.62$ while including
the data with $q=2$ gives $\sigma_{z_{spec}}=0.42$ and
$\sigma_{z_{opt}}=0.53$. Bootstrap resampling of the true
(spectroscopic) distribution indicates that the widths of the
inversion and spectroscopic distributions are not statistically
distinguishable at greater than $\sim75\%$ confidence. We do expect,
however, that the uncertainties in the inversion redshifts will
increase the predicted dispersion over the spectroscopic one. We may
therefore consider the inversion-predicted dispersion better as an
upper limit on the true width of the redshift distribution.  In
summary then, this analysis gives us confidence that our method can
find the correct mean redshift for an unknown distribution, but that
the width of the distribution is less well determined. For the mean
redshift, the agreement between the predicted and spectroscopic values
is especially encouraging when one considers the small sample size.

\section{Implications for the Mass Model}

In this section we return to the mass model used by KESCS and investigate
to what extent the new spectroscopic results further constrain it. 

First, we consider only the spectroscopic redshifts of the multiple
images, namely: \#359 (z=0.702), \#259 (z=1.034), \#384 (z=2.515), and
\#444 (z=1.030) and optimize the different cluster and galaxy scale
components following the KESCS prescription.  Following the analysis of
Natarajan \& Kneib (1997), we also vary the exponent $\alpha$ of the
scaling law of the truncation radius ($r_{cut} = r_{cut}^*
(L/L_*)^{1/\alpha}$).  We tried two different values: $\alpha=4$ to
mimic a constant mass to light ratio for the galaxies, and
$\alpha=2.5$ to mimic a `Kormendy' relation between the effective
radius of the mass and the luminosity.  The resulting best fit mass
distribution does not differ greatly from the one presented in
KESCS. Less mass is required between the two main clumps than in the
KESCS model, and the $\alpha=2.5$ exponent is preferred leading to a
mass-to-light ratio of $9$ for an $L_*$ galaxy.

The improved model in turn allows us to put better constraints on the
redshifts of the other multiple images.  In particular, the \#H1-3
multiple arc appears to be clearly different from the arclet \#273
(z=0.80) as its redshift is now predicted to be $z=1.9\pm 0.2$.  In
fact, closer inspection of the HST/WFPC2 image reveals a small offset
between the \#273 and \#H1-3 systems of $\sim$1 pixel, suggesting that
these images are indeed separate systems. The multiple system \#730
suffers the same translation to higher z as {\bf H1,2,3} and is now
expected to be at a higher redshift than $\sim$3. (With the relatively
low redshift of A2218, our ability to state tighter constraints for
high redshift arcs is hindered by the weak dependance on redshift of
the $D_{LS}/D_S$ ratio.)

We then considered both the singly and multiply imaged arclets whose
spectroscopic redshifts are presented here. Their images were mapped
back to the source plane to compare their intrinsic shape distribution
to that of the unlensed field (shown by contours in Figure
\ref{fig-plottau}). These intrinsic ellipticities form a circularly
symmetric distribution in the $\bar{\tau_S}$-plane and, with few
exceptions, have ellipticities within $\tau<0.2$. This area contains
over half of the entire unlensed distribution and therefore the predicted
shapes of our arclets well match those expected for unlensed field
galaxies.  To include the single image redshifts in the optimisation,
we reran the optimisation procedure, adding to the $\chi^2$ sum a term
of the form $(\tau_S/\sigma_{\tau S})^2$ for each arclet, where
$\sigma_{\tau S}$ is the width of the intrinsic ellipticity
distribution.  The difference between the two optimizations is minor,
demonstrating that we are now very close to the true mass distribution
within the boundary of the HST/WFPC2 field.

\begin{figure}
\psfig{file=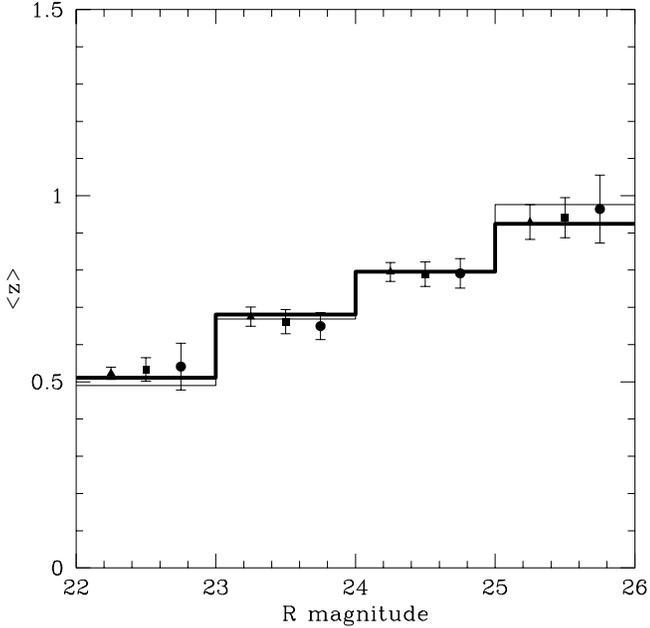,width=0.5\textwidth,angle=0}

\caption{The effect on the redshift predictions of removing galaxy
scale clumps from the model . Points with error bars show the mean and
dispersion in $<\! z\! >$ for each bin. {\it triangles:} One clump removed
at a time. {\it squares:} 50 random pairs of clumps removed one at a
time. {\it circles:} 50 random sets of 5 clumps removed one at a
time. The three sets of points are offset in R for clarity. In each
case, the thin line shows the mean redshift--magnitude relation for
the fiducial model (KESCS) while the thick solid line shows the
relation predicted by our new model developed in \S4. }

\label{fig-removegal}
\end{figure}

\begin{figure}
\psfig{file=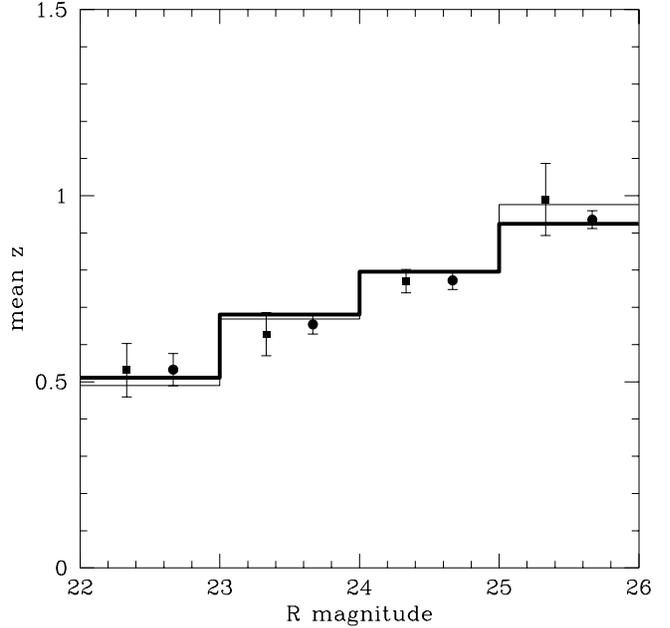,width=0.5\textwidth,angle=0}

\caption{The effect of adding large scale external mass concentrations
to the model ({\it squares}) and of varying the cut radius of the main
cluster clump ({\it circles}). Each point and error bar shows the the
mean and dispersion in the $<\! z\! >$ for each bin and the points are
separated in R for clarity as before. Again the thin solid line
represents the fiducial model of KESCS while the thick line shows the
predictions from our new model of \S4. }
					
\label{fig-addcl_rcut}
\end{figure}

Finally, we used the updated model on the catalogue of faint arclets
with areas above 50 pixels (as used in KESCS) to compute the mean
redshift versus magnitude prediction. This is shown by the thick line
on Figures \ref{fig-removegal} and \ref{fig-addcl_rcut} where we
compare it to the prediction of KESCS (thin lines). Clearly, while
there may be some differences in $z_{opt}$ for individual arclets, the
difference in the prediction for the whole sample (even in the
faintest bin) is small ($\ltapprox5\%$).  The new mass model is the
best we can achieve with the present redshift constraints.

We now discuss how the residual uncertainties in the mass model can
affect our redshift predictions. Of course, we can only change the
mass model in such a way that it still reproduces the multiple imaging
seen in the HST image. In the modelling process (see KESCS) we use a
$\chi^2$ sum to represent how well the predicted parameters of the
multiple images match those observed. Thus, in perturbing our model,
we will try to keep the resulting $\chi^2$ as close to that of the
fiducial model as possible. A number of tests may be made,
corresponding to uncertainties in the model on both small and large
scales.

Firstly, we may investigate the effect of the galaxy scale
components. These were included in the model solely to reproduce
several of the multiple image systems correctly. Only a few were
actually needed to do this but a larger number ($\sim30$) were added
corresponding to a given magnitude limit in the interests of
consistency. We tested their effect on the redshift predictions by
removing them at random from the model and predicting the redshift -
magnitude relations, as shown in Figure \ref{fig-removegal}. The
triangles, circles and squares show the mean and dispersion in $<\!
z\! >$ for each bin when clumps were removed in groups of 1, 2 and 5
respectively. We also show the predictions of our new model (thick
line) and that published by KESCS (thin line) where both include the
statistical correction for cluster and foreground contamination as
described in KESCS. The only individual cases deviating significantly
from the fiducial predictions were those where the galaxy scale clump
corresponding to the cD galaxy was removed. However, since an external
constraint in the form of the cD velocity dispersion exists for this
clump, we do not consider its removal to be a valid test and exclude
these cases from the analysis. In all three cases, the dispersion in
$<\! z\! >$ for each bin never rises above $\sigma_{<\! z\! >}=0.09$
and the average fractional dispersion is only $\sigma_{<\! z\! >}/<\!
z\! >=5.6\%$. Clearly, the mean redshift and its trend with magnitude
are insensitive to the details of the mass model as represented by the
galaxy scale clumps.

On large scales ($\sim1Mpc$), the uncertainties in the mass model
derive mainly from the small angular size of the HST image. Outside
the region of multiple imaging, we obtain less information on the mass
profile, and features such as the extent of the main cluster clump or
the presence of external mass clumps are poorly constrained. However
the effect of these uncertainties can be readily tested in a similar
manner to that of the galaxy scale clumps. Figure \ref{fig-addcl_rcut}
shows the effect on the magnitude--redshift relation of two such
tests: adding external large scale clumps outside the HST area
(squares) and varying the cut radius of the main cluster clump
(circles). Again, each point shows the mean and dispersion in $<\! z\!
>$ for each bin, and we show the predictions of KESCS and our new
model by thin and thick lines respectively. For the first test, large
scale mass clumps varying in velocity dispersion ($500-2000km/s$) and
core radii ($700-1500kpc$)\footnote{Here and in the following we
assume $H_o = 50$ km s$^{-1}$ Mpc$^{-1}$ and $q _o=0.5$} were added to
the model at the same redshift as the cluster and four different
positions $1.9$ Mpc from the centre of the cluster (outside the HST
frame). 48 different combinations of clump parameters were tried and
in each case the model was re-optimized in order to reproduce the
multiple image constraints as well as possible. Only the models giving
a residual $\chi^2$ within 30\% of the fiducial model value were
considered. Again, the maximum dispersion is low, $\sigma_{<\! z\!
>}=0.10$ and the average fractional dispersion is just $\sigma_{<\!
z\! >}/<\! z\!  >=6.4\%$. For the second test, the cut radius of the
main cluster clump, $r_{cut}$, was varied between $400-2000kpc$,
re-optimizing the model to reproduce the multiple images as
before. Here the maximum dispersion is $\sigma_{<\! z\! >}=0.04$ while
the average fractional dispersion is $\sigma_{<\! z\! >}/<\! z\!
>=4.5\%$. Once more, we find that the uncertain features of our model
on large scales do not affect the results appreciably. In summary
then, our redshift predictions for the faint population behind A2218
are unlikely to be much affected by the remaining uncertainties in our
mass model.

\section{Properties of the Faint Population}

The results of our spectroscopic program presented in section \S3 give
strong support to the inverted redshift distribution derived for
fainter arclets viewed through Abell 2218. Moreover, the discussion of
\S4 demonstrates that further changes to the mass model permitted by
our spectroscopic data will not greatly affect these redshift predictions.

For arclets with well measured shapes, such as our spectroscopic
sample, we found in \S3 an error in the {\it mean} redshift of
$\sim2\%$. Using fainter arclets with more poorly measured shapes
would increase the errors for each individual arclet and thus increase
the uncertainty in the mean redshift for the population as a
whole. This point was addressed fully in KESCS where it was shown
using simulations that reliable shapes could be derived for arclets
with isophotal areas above 50 pixels. We must therefore restrict
ourselves to this area limit in any analysis of the population. The
results (Figures 15--16) show a mean redshift $<\! z\! >\simeq$0.8-1.0
for amenable sources selected with $R\simeq$25.5 ($B\sim 26$--27).

A low mean redshift at such a faint limit has consistently emerged
from statistical lensing studies (Smail et al.\ 1994, KESCS). Recently,
Luppino \& Kaiser (1997) have reported the detection of a significant
weak shear signal from the bluest half of the faint field population
behind a distant cluster, inferring a high mean redshift ($z>>1$) for
this subsample.  However, we note that by combining the blue and red
subsets from Luppino \& Kaiser's analysis the mean redshift they would
infer for the {\it whole} faint galaxy population would be much closer
to the no evolution expectation (in line with the results presented
here) especially when allowance is made for the various uncertainties
in their measurements and analysis.

The only independent technique which has been used
to investigate the redshift distribution of galaxies beyond the
spectroscopic limit is that based on multicolour photometry (the
photometric redshift technique). The popularity of this technique has
increased considerably of late because of precise data available for
the Hubble Deep Field (HDF). It is thus of interest to compare our
results with those of workers who have determined photometric
redshifts in the HDF. Several surveys have been published using the
HDF optical photometry combined with spectral templates derived either
from population synthesis or local spectral energy distributions or
both.  The advantages of adding near infrared photometry to follow the
usual spectral features to redshifts beyond $\sim1.3$ are clear, but
are difficult to compare with our results since they are usually
magnitude limited in some infrared band. Here we compare our redshift
distribution with the results of Mobasher et al.\ (1996) and Lanzetta et
al. (1996).

\begin{figure}
\psfig{file=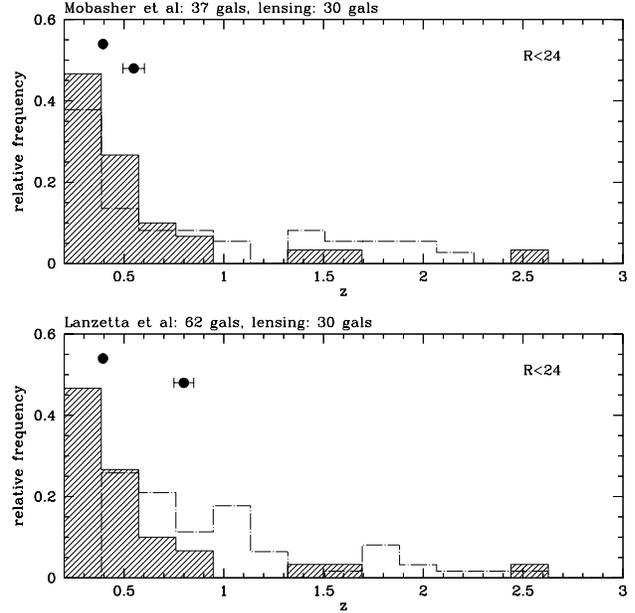,width=0.5\textwidth,angle=0}
\caption[]{Comparison of the lensing $n(z)$ (shaded) with the
photometric $n(z)$ from the Hubble Deep Field analyses of Mobasher et
al ({\it top}) and Lanzetta et al.\ ({\it bottom}) for the magnitude
limit $R<24$. Points with errorbars show the median of each
distribution, with errors calculated from bootstrap resampling. In
order to make a valid comparison, only the largest 59\% of objects
were used. It is clear that there is much closer agreement with
Mobasher et al than Lanzetta et al, emphasizing the discrepancies
between the two photometric n(z) determinations.}
\label{fig-nz24}
\end{figure}

Such comparisons are, however, hindered by the different techniques
used and, in particular, that it is difficult to construct a magnitude
limited sample from the lensing inversion method. Firstly, since the
magnification of images varies across the cluster, our observed
magnitude limit does not correspond to a sharp intrinsic magnitude
limit, rather the lensing sample suffers from a more gradual fall off
in counts with magnitude. One approach would therefore be to choose an
effective magnitude limit corresponding to the point where the counts
fall significantly below those in the field. The second complication
arises from the area limit discussed above. In this case, we can
choose to select the same proportion of faint galaxies as a function
of area from both the lensing and photometric redshift samples. These
choices correspond to first making a magnitude cut at $R\simeq24$
($I\simeq23.5$) and then taking only the largest $59\%$ of the
sample at this limit. Figure \ref{fig-nz24} shows the comparisons,
where we have chosen to use a slightly deeper magnitude limit ($I<24$)
for the photometric redshift distributions in order to obtain a
comparable number of galaxies from the HDF.

\begin{figure}
\psfig{file=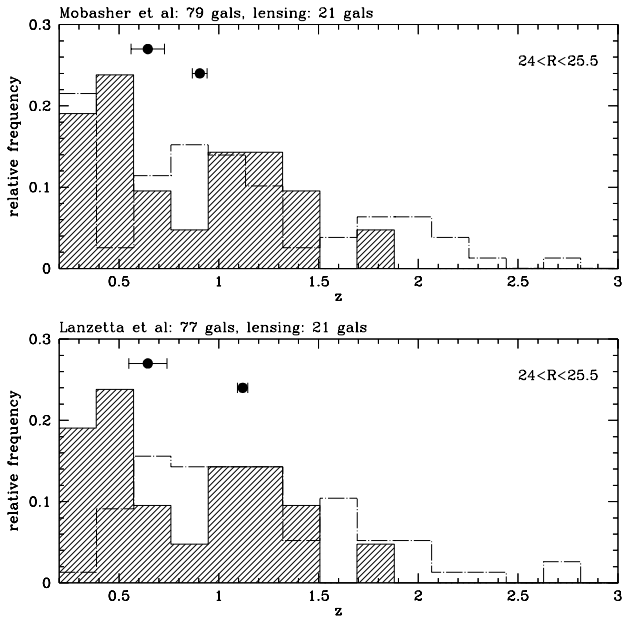,width=0.5\textwidth,angle=0}
\caption[]{Comparison of the lensing $n(z)$ (shaded) with the
photometric $n(z)$ of Mobasher et al.\ ({\it top}) and Lanzetta et al
({\it bottom}) for the magnitude slice $24<R<25.5$
($23.5<I<25$). Again, points with errorbars show the median of each
distribution, and only the largest 59\% of objects used. The lens
inversion method predicts a significantly lower median redshift than
those based on {\it optical} multi-colour photometry at these
limits. }
\label{fig-nz24_25.5}
\end{figure}

It can readily be seen that the agreement is much better when we
compare to the results of Mobasher et al. than Lanzetta et
al. (although the numbers of galaxies are small.) Excluding the
galaxies below $z=0.2$ (which cannot be probed by our lensing
analysis), we find the lensing n(z) has a median of $z=0.4$ with only
7\% of objects appearing above $z=1.5$ while Mobasher et al.'s
distribution has a slightly higher median redshift ($z=0.5$) but with
a longer tail (19\% above $z=1.5$). The corresponding numbers for the
Lanzetta et al. distribution are $z=0.8$ and 18\%. However, as one
goes fainter, the agreement worsens. Figure \ref{fig-nz24_25.5} shows
the predicted redshift distributions for our faintest magnitude slice
$24<R<25.5$. At this limit, the lensing method predicts a low median
redshift of $z=0.7$ while the corresponding values from optical
multi-colour photometry are $z=0.9$ (Mobasher et al.) and $z=1.1$
(Lanzetta et al.) respectively.
 
At the faintest limits, it is important to bear in mind that the
sensitivity of photometric redshifts in the $1<z<2$ region is expected
to be very poor as there are no significant continuum features at
optical wavelengths. For $z>$1.3, the 4000 \AA\ break disappears
from the optical window and the Lyman break does not enter until
$z>2$. Is it therefore possible that some galaxies allocated to
$1.5<z<2.0$ by the photometric technique may actually lie at lower
redshifts ?

An important recent development in this regard is the addition of JHK
ground-based near-infrared data to complement the earlier optical
photometric redshift studies. Connolly et al.\ (in prep.) report a
redshift distribution based on 4-colour optical plus 3-colour near
infrared data for a $J<$23.5 sample which has $<\! z\! >$=0.9$\pm$0.1. This
corresponds to a limit somewhat deeper than $R=25.5$ but is a good
indication of the changes that we might expect when near-infrared data
is added to the studies discussed above.
 
\section{Conclusions}

We summarize our conclusions as follows:

\begin{itemize}

\item We have conducted one of the deepest spectroscopic surveys
attempted on a 4m class telescope and certainly the faintest with
LDSS-2. The faintest redshifts were obtained for sources with $R$=24 whose
unlensed magnitudes reach $R$=25 in some cases. The major limiting
factor is the effectiveness of sky subtraction and our limit is more
appropriately constrained by surface brightness than integrated
magnitude. We have achieved a useful success rate to $\mu_R=24$
arcsec$^{-2}$ with completeness falling below 50\% at $\mu_R=23$
arcsec$^{-2}$.

\item The results of our spectroscopic survey show that the lensing
inversion technique is reliable. The mean redshift of our spectroscopic
sample determined from both inversion and spectroscopic redshifts
agree to within $\sim2\%$.  Since the inversion method is purely
geometric in operation, this agreement gives us confidence when
estimating the mean redshift of the larger population of faint arclets
provided their individual shapes can be adequately measured.

\item We have included the new spectroscopic redshifts to further
constrain our mass model and find that the redshift predictions change
by a maximum of $\sim5\%$. Moreover, the predicted redshifts for a
population of arclets with $R\simeq$25.5 are insensitive to further changes
in the mass model allowed by the multiple image constraints. We
conclude that errors in the inferred mean redshift deriving from
uncertainties in the mass distribution can remain only below the level
of $\delta_{<\! z\! >}/<\! z\! > \ltapprox 10\%$.

\item The new predictions for the faint population do not differ
significantly from those presented by KESCS. The mean redshift for the
population of arclets amenable to inversion with $R\simeq$25.5 is broadly
consistent with the no evolution prediction; we find $<\! z\! >\simeq0.8-1.0
$.

\item To a limit set by our magnitude and area cuts, the inversion
prediction agrees reasonably well with results presented by other
authors using photometric redshifts determined from data on the Hubble
Deep Field. A direct comparison is difficult since the inversion
technique requires an area limited survey in contrast to one that is
magnitude limited. The agreement may be much better when
near-infrared photometry is used in addition to optical colours. This
can be understood via a greater sensitivity to the 4000 \AA\ break
which is redshifted into the near-infrared at redshifts of interest.

\end{itemize}

Application of this technique to other clusters with similar data has
already begun (Kneib et al. in prep.). This will generate a much
larger sample of inverted arclets and hence beat down the statistical
uncertainties, as well as overcoming the effects of redshift
clustering in the background population. Using cluster lenses over a
wide range of redshifts will reduce the intrinsic uncertainties coming
from the redshift dependence of the distance ratio, while extending
the technique into the near-infrared using NICMOS will allow us to
select $1<z<3$ galaxies in their restframe optical, to provide a more
robust measure of the star-formation density at these epochs. By
expanding our data set in this way, we shall be able to use this {\it
proven} technique to derive robust and useful parameters for the faint
population of galaxies at depths presently unobtainable by
conventional means.

\section{Acknowledgments}

We thank Mike Breare and all the La Palma support staff for ensuring
the smooth operation of LDSS-2 on the WHT. We especially thank Karl
Glazebrook for the use of his {\it REDSHIFT} utility. We thank Nial
Tanvir and Karl Glazebrook for their expertise with both LDSS-2 and
the LEXT software and we thank Bernhard Geiger for useful
discussions on the inversion method. TMDE and IRS acknowledge financial
support from PPARC.

\end{document}